\documentclass{sig-alternate}
\pdfoutput=1
\newcommand{\ignore}[1]{}
\usepackage[pass]{geometry}
\usepackage{fancyhdr}
\usepackage[normalem]{ulem}
\usepackage[hyphens]{url}
\usepackage{hyperref}
\usepackage{color}
\usepackage{soul}
\usepackage{caption}
\usepackage{subcaption}
\usepackage{algorithm2e}

\def\sharedaffiliation{%
\end{tabular}
\begin{tabular}{c}}

\title{IMPACT: Interval-based Multi-pass \\Proteomic Alignment with Constant Traceback}

\begin{document}

\numberofauthors{3}
\author{
    \alignauthor Sahand Kashani \\
    \email{sahand.kashani@epfl.ch}
    \alignauthor Stuart Byma \\
    \email{stuart.byma@epfl.ch}
    \alignauthor James R. Larus \\
    \email{james.larus@epfl.ch}
    \sharedaffiliation
    \affaddr{Very Large Scale Computing Laboratory (VLSC)} \\
    \affaddr{EPFL}   \\
    \affaddr{Switzerland}
    }

\maketitle


\begin{abstract}

    Darwin~\cite{Turakhia:2018:DGC:3173162.3173193} is a genomics co-processor that achieved a $15000\times$ acceleration on long read assembly through innovative hardware and algorithm co-design.
    Darwin's algorithms and hardware implementation were specifically designed for DNA analysis pipelines.
    This paper analyzes the feasibility of applying Darwin's algorithms to the problem of protein sequence alignment.
    In addition to a behavioral analysis of Darwin when aligning proteins, we propose an algorithmic improvement to Darwin's alignment algorithm, GACT, in the form of a multi-pass variant that increases its accuracy on protein sequence alignment.
    Concretely, our proposed multi-pass variant of GACT achieves on average 14\% better alignment scores.
\end{abstract}

\section{Introduction}
\label{sec:introduction}

Biological sequences are compared through sequence alignment\footnote{We use this term to refer to dynamic programming ``exact'' alignment~\cite{Smith:1981:IPA:0022-2836(81)90087-5} between two sequences, rather than mapping of reads to a reference, which is sometimes also called \textit{alignment}.}:
Given two strings of different lengths, called the query $Q$ and the reference $R$, alignment consists of finding the location of a series of gaps to insert into both sequences to make them align, subject to a scoring function that rewards matches and penalizes mismatches or gaps.

Smith \& Waterman (S\&W) proposed a dynamic programming algorithm capable of finding an \emph{optimal} alignment in time and space $O(|Q| \times |R|)$ \cite{Smith:1981:IPA:0022-2836(81)90087-5}.
The algorithm has two phases: (1)~computing a $|Q| \times |R|$ scoring matrix representing the alignment scores of all possible sub-sequences of $Q$ and $R$, and (2)~tracing back from the highest-scoring cell in the matrix through a combination of 3 directions ($\leftarrow, \nwarrow, \uparrow$) until a cell with score zero is found.
The sequence of cells encountered during the traceback forms the optimal alignment of the two sequences.
The S\&W algorithm is compute-intensive and has been the focus of many acceleration efforts such as parallelization~\cite{farrar2006striped, swps3} and hardware acceleration~\cite{smithwatermanfpga, smithwatermansystolic}.

Recent efforts include Darwin~\cite{Turakhia:2018:DGC:3173162.3173193}, a system that accelerates alignment through both hardware and algorithmic innovations.
Crucially, Darwin provides mechanisms for hardware-accelerated alignment with traceback support (providing not only the alignment score, but also the sequence of edits through which the score was achieved) using a \emph{constant} amount of memory.
This allows Darwin to align arbitrary-length sequences using space-limited hardware resources, achieving orders of magnitude speedups on reference guided and de-novo assembly operations.
Darwin uses D-SOFT (Diagonal-band Seed Overlapping based Filtration Technique) to reduce the dynamic programming search space, followed by GACT (Genome Alignment using Constant memory Traceback) to perform the alignment.
D-SOFT finds good exact-matching \emph{seeds} or subsequences between the two sequences being aligned, which GACT then uses as its starting point to perform the \emph{extension} of the alignment, extending from both sides of the seed until an alignment is found.
Both component algorithms of Darwin are heuristic, and while Darwin is not guaranteed to find the optimal alignment, it does so in practice most of the time.

Darwin's algorithms were designed and evaluated with DNA sequences, but fast alignment is also needed for protein sequence analysis.
Proteins are different from DNA --- they consist of a much larger alphabet, they have highly variable lengths, and their alignment substitution matrices are non-uniform (that is, each alphabet character pair is not a simple match or mismatch, but rather encodes a substitution probability).

These differences lead to some surprising and sometimes counter-intuitive behavior from Darwin that results in suboptimal performance when applied as-is to protein sequences.
In this paper, we provide an analysis of this behavior and investigate the underlying causes.
In addition, we propose a new multi-pass version of GACT that helps improve alignment scores for proteins.
More specifically, this paper makes the following contributions:
\clearpage
\begin{itemize}
\item[--] A detailed analysis of the performance of Darwin when used to align protein sequences.
\item[--] Based on this analysis, a proposal and evaluation of an algorithmic augmentation to GACT to retain near-optimal alignment results for proteins.
\end{itemize}

The rest of this paper is organized as follows:
Section~\ref{sec:gact_aligner} describes the GACT algorithm;
Section~\ref{sec:aligning_proteins_with_gact} discusses situations where GACT fails to provide optimal alignments for protein sequences;
Section~\ref{sec:multi_pass_gact} introduces a multi-pass version of GACT that achieves near-optimal alignment of ``straight-line'' sequences;
Section~\ref{sec:interval_alignment} describes future work related to joining intermediate multi-pass GACT results together;
Section~\ref{sec:conclusion} concludes.

\section{The GACT aligner}
\label{sec:gact_aligner}

GACT takes a candidate seed produced by D-SOFT and left- and right-extends it in a \emph{tiled} fashion using the S\&W algorithm.
To simplify the following description, our discussion assumes GACT is seeded with the max-scoring cell of the S\&W scoring matrix and that only left-extension is performed.
GACT has two novelties compared to prior hardware S\&W implementations:

\begin{itemize}
    \item[--] GACT computes a \emph{single} tile of the S\&W matrix and traces it back \emph{before} moving to the next tile, whereas other implementations only start a traceback \emph{after} computing the \emph{complete} S\&W matrix.
    The advantage is that the amount of traceback state to be stored in on-chip memory is limited by the tile size $T$.
    \item[--] GACT does not trace back complete tiles, but rather stops at a distance $O$ from one's sides and uses this position as the starting offset of the next tile to be computed.
    The end result is that neighboring tiles \emph{overlap} and some cells are computed twice, possibly to \emph{different} values.
\end{itemize}

\begin{figure*}
    \centering
    \begin{subfigure}{.33\textwidth}
        \centering
        \includegraphics[width=1.0\linewidth]{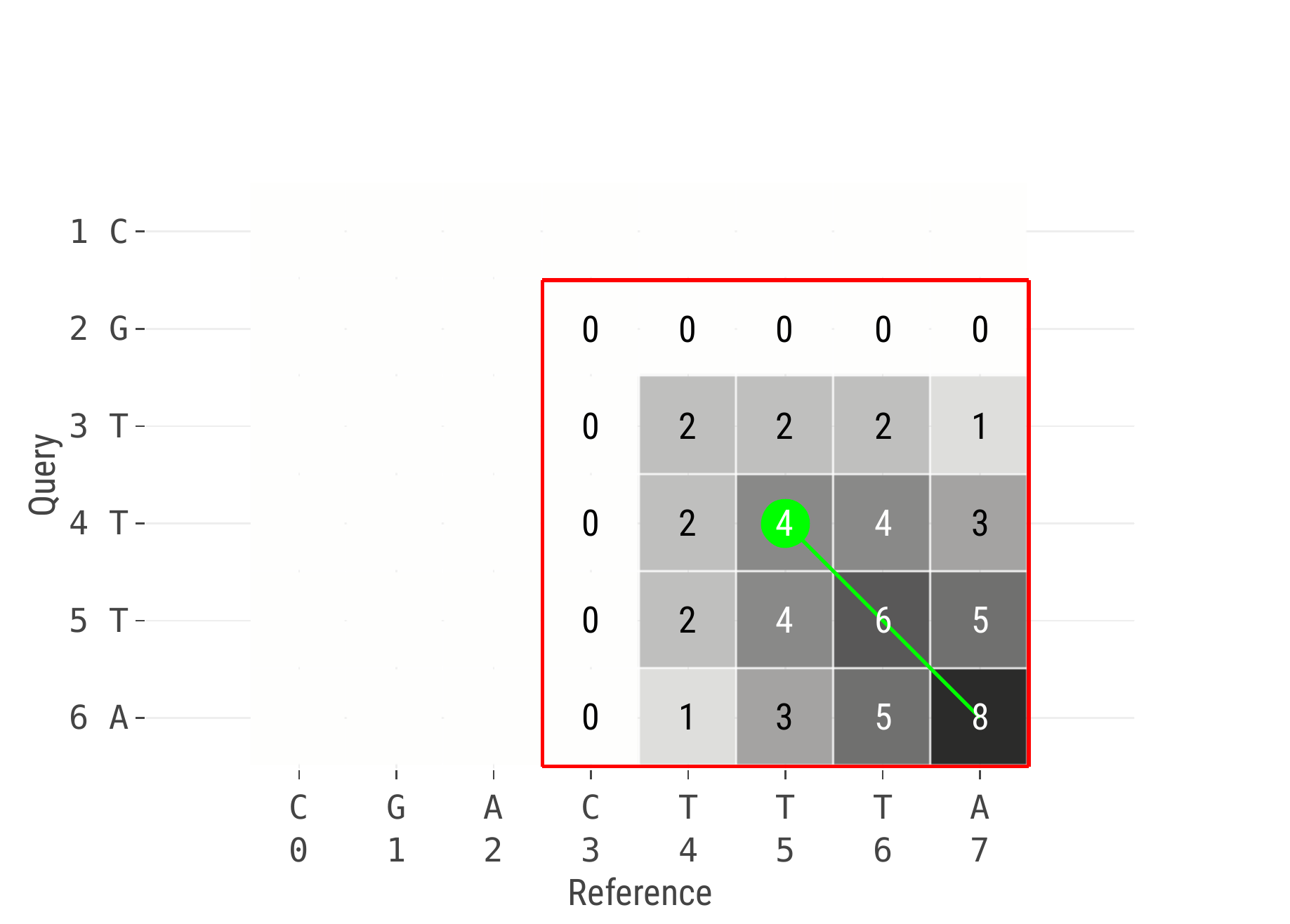}
    \end{subfigure}%
    \begin{subfigure}{.33\textwidth}
        \centering
        \includegraphics[width=1.0\linewidth]{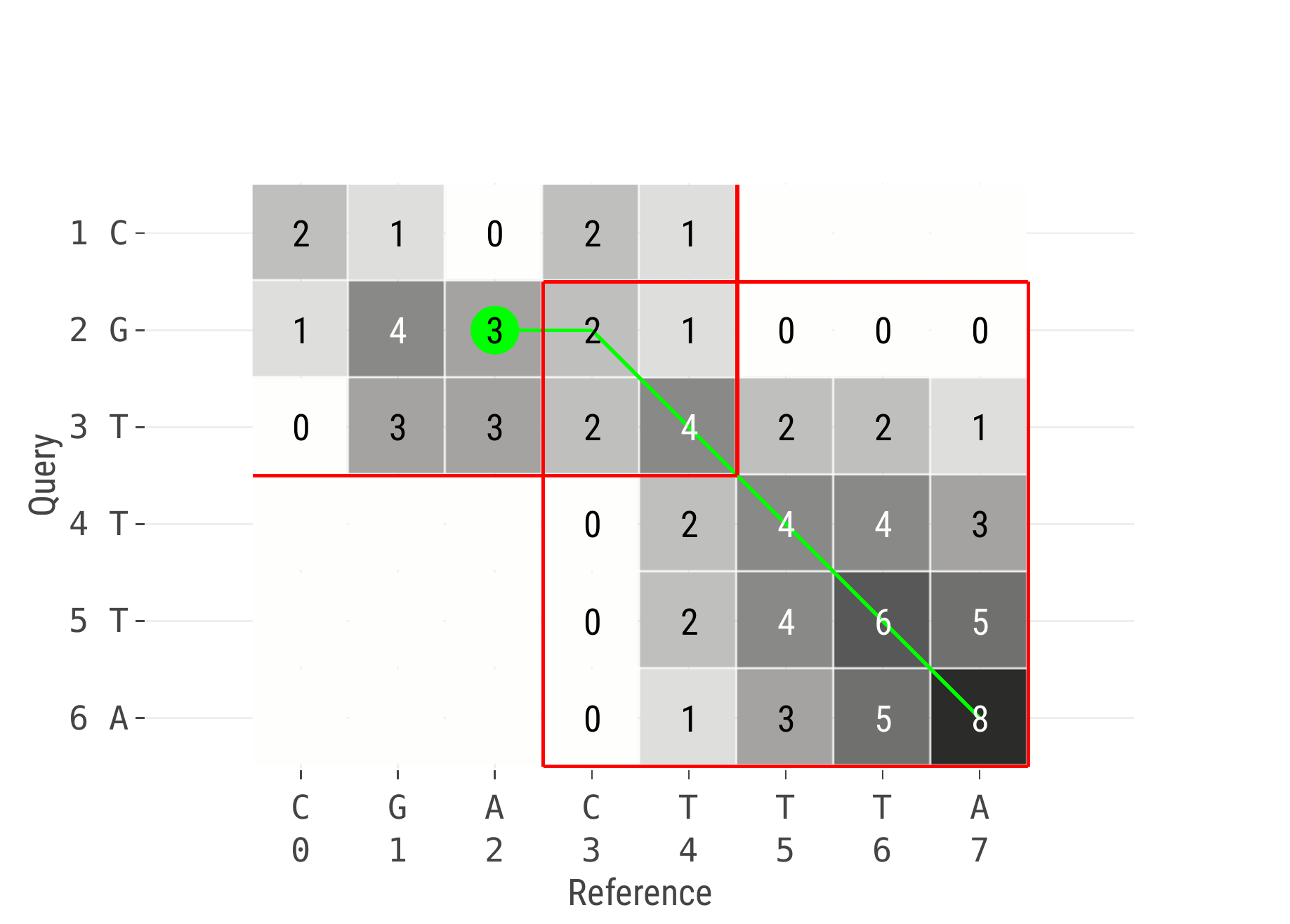}
    \end{subfigure}%
    \begin{subfigure}{.33\textwidth}
        \centering
        \includegraphics[width=1.0\linewidth]{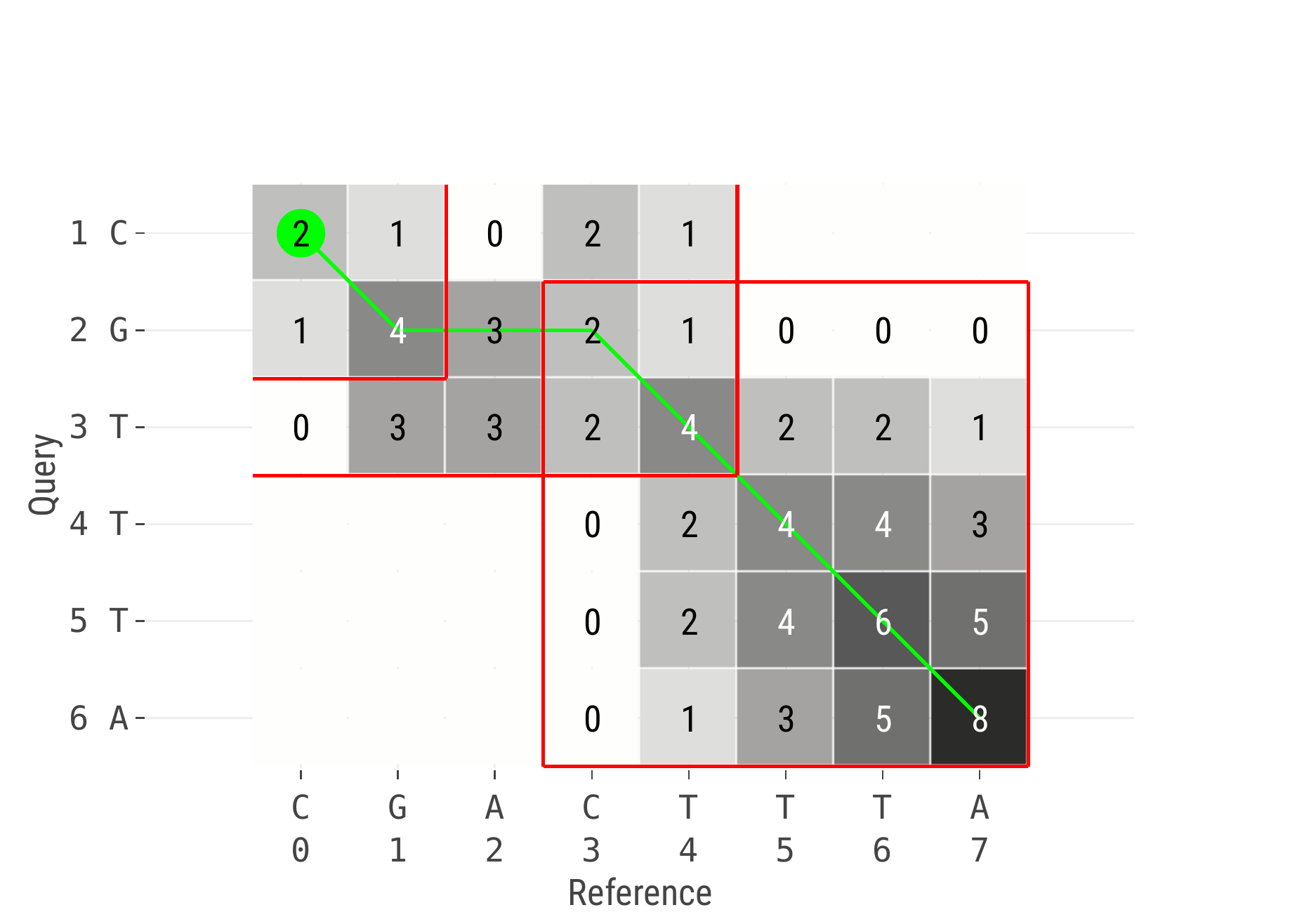}
    \end{subfigure}
    \caption{GACT with tile size $T=5$ and overlap $O=2$ performing the left extension of a seed-and-extend algorithm.
    In this example, matches are rewarded +2 and mismatches/gaps are penalized by -1.
    Notice how the values of overlapping cells increase between successive tiles.}
    \label{fig:gact_illustration}
\end{figure*}

GACT is \emph{not} an optimal alignment algorithm since the value of a cell, as defined by S\&W, depends on its 3 neighbors ($\leftarrow, \nwarrow, \uparrow$), but these neighbors do not exist for the top and left borders of a tile.
However, GACT's use of overlapping tiles enables recomputed cells to obtain \emph{higher} values than those computed in their first tile\footnote{Cells in the overlapping region of the first tile are close to the tile borders and are therefore very sensitive to border values. The same cells in the second tile are far from the borders and are less sensitive to its values.}.
Overlapping tiles therefore allow a traceback to follow a path that would otherwise not have existed if a cell value was not recomputed.
This behavior can be seen at the interfaces between the tiles in Figure~\ref{fig:gact_illustration}.

GACT uses the unmodified S\&W algorithm in each tile.
It is efficiently implementable in hardware using a systolic architecture of processing elements (PEs) that computes tile cells in a wavefront fashion.

\section{Aligning Proteins with GACT}
\label{sec:aligning_proteins_with_gact}

\begin{figure}[h]
    \centering
    \includegraphics[width=0.65\columnwidth]{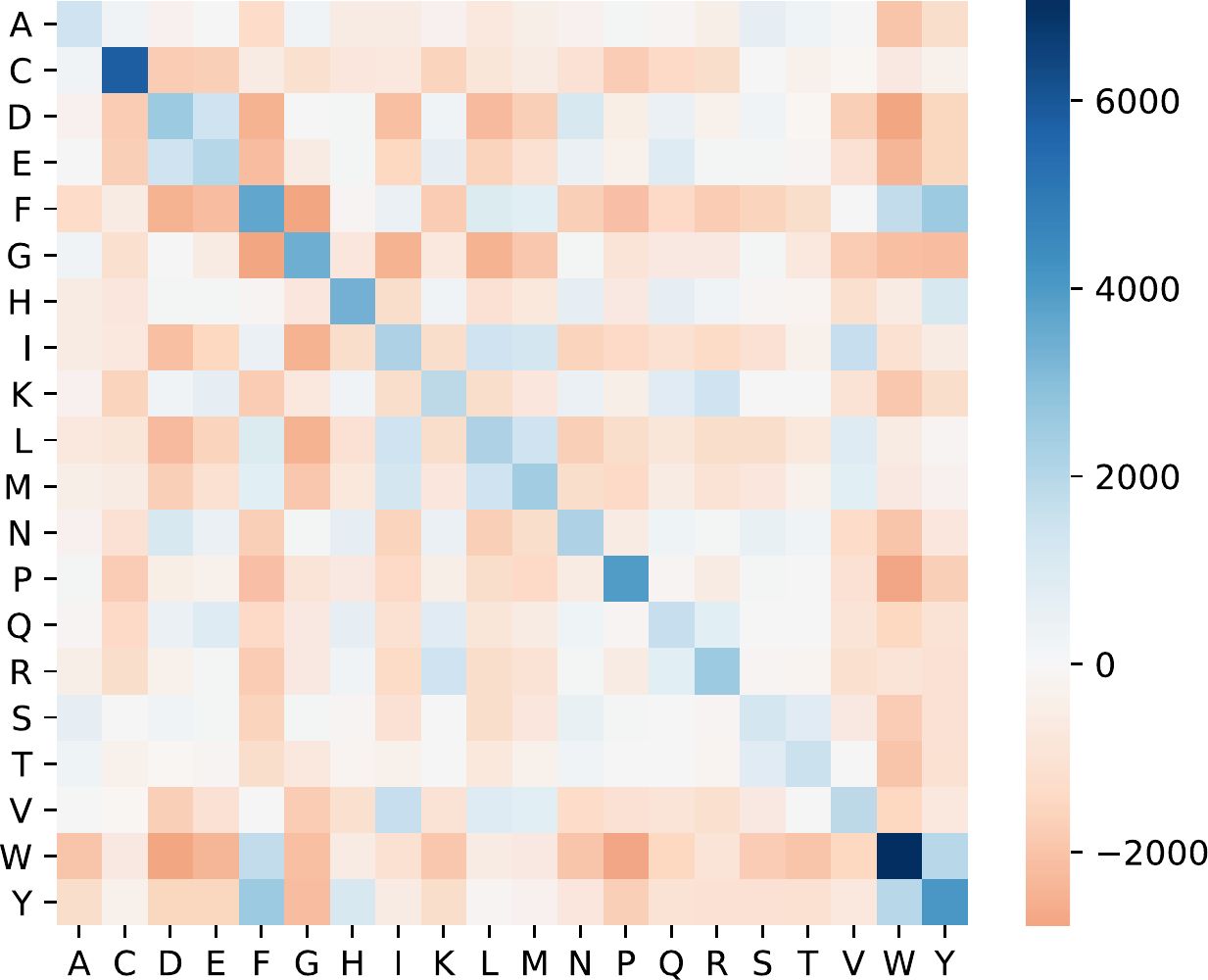}
    \caption{Sample protein substitution matrix. Note how some amino-acid matches are rewarded much higher than others (C-C, W-W) and that certain mismatches are also awarded positive substitution scores.}
    \label{fig:protein_substitution_matrix}
\end{figure}

The recurrence relation in the S\&W algorithm involves a table lookup into a \emph{substitution matrix}.
This symmetric matrix determines the weight attributed to each character pair in the S\&W algorithm and is the primary reason for which GACT does not always achieve optimal alignments on protein sequences.
There are several important differences between DNA and protein substitution matrices.
First, while DNA is formed from nucleotides with an alphabet size of just 4 characters: \{A, C, G, T\}, proteins are formed from sequences of amino-acids, building blocks with a much larger alphabet size of 20, leading to much larger matrices.
Second, DNA sequences have simple matrices that attribute \emph{homogeneous} scores to all matches and mismatches (e.g.~+2 for matches, -1 for mismatches), whereas protein matrices are \emph{heterogeneous} and weigh not only matches positively, but also certain mismatches.
Finally, entries in protein substitution matrices vary wildly in magnitude, with certain matching amino-acid pairs weighing up to $5\times$ more than others (c.f.~Figure~\ref{fig:protein_substitution_matrix}.)

Put together, these differences make it difficult for GACT to converge back to the path taken by the optimal protein alignment when it makes an erroneous local traceback decision.

\subsection{More tile overlaps are not always better}
Darwin showed that (1) increasing $T$ while keeping $O$ constant, and (2) increasing $O$ while keeping $T$ constant both result in an increase in the number of observed alignments that are optimal (with best results obtained when both components are increased).

Darwin's observations, however, do not always hold for protein sequences.
Consider the following experiment: Given a $(T, O)$ configuration for which there exists some proteins that do not align optimally, proceed to incrementally increase $O$ (while keeping $T$ constant) until the number of proteins that align optimally increases.
Intuitively, one expects that increasing $O$ will always increase the number of relatively-high scoring cells in an overlapped region, and therefore provide more confident values for the traceback to use.
So increasing $O$ should enable GACT to find the same or better alignments than at lower overlap margins.
However, we actually observe that the total number of optimal alignments obtained by GACT by following this procedure \emph{decreases}!
This implies some of the alignments that were previously optimal with a smaller value of $O$ now diverge with larger values.

\begin{figure*}
    \centering
    \begin{subfigure}{.45\textwidth}
        \centering
        \includegraphics[width=0.775\linewidth]{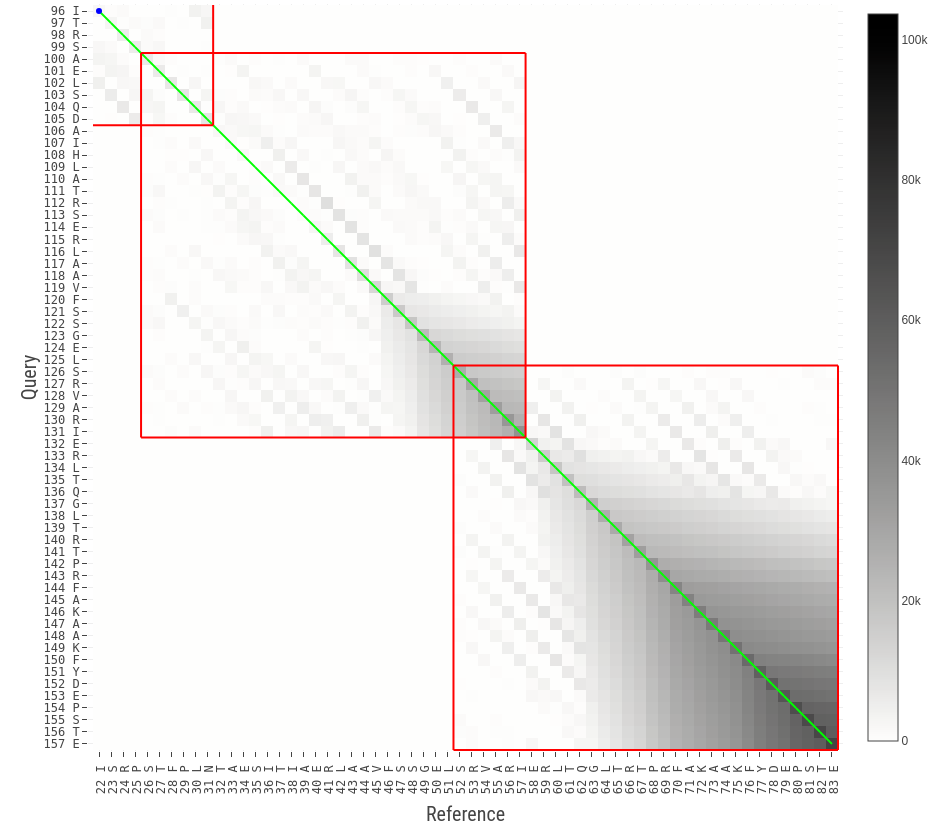}
        \caption{GACT alignment with $(T, O) = (32, 6)$.}
        \label{fig:single_pass_more_overlap_not_always_better_small}
    \end{subfigure}%
    \begin{subfigure}{.45\textwidth}
        \centering
        \includegraphics[width=0.775\linewidth]{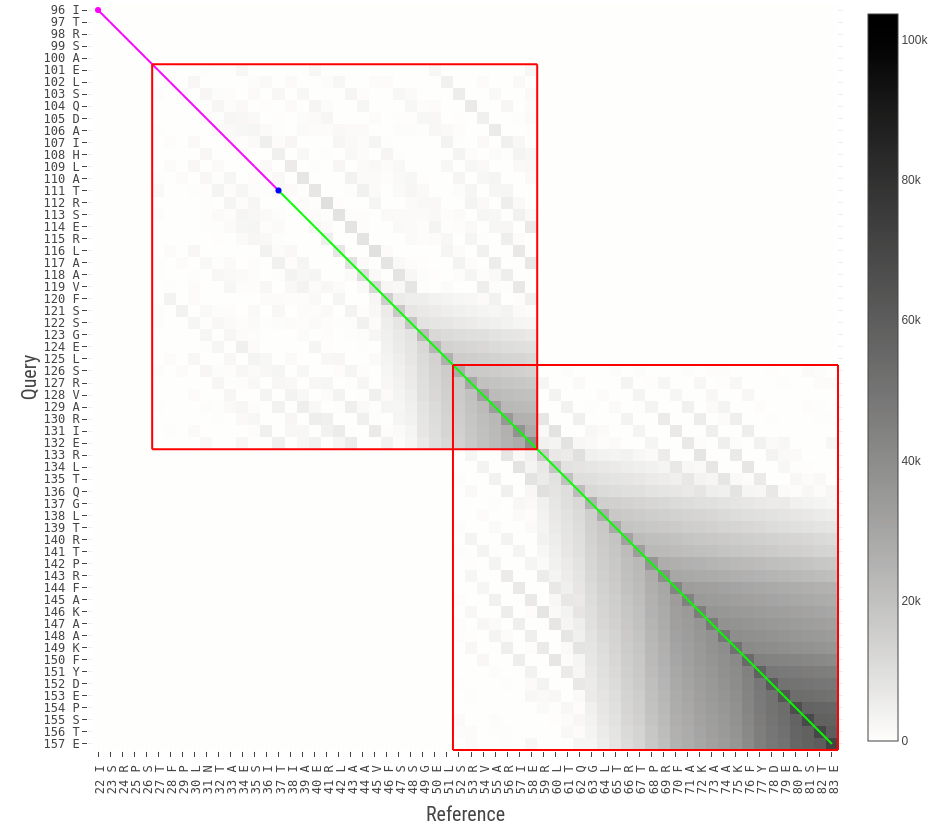}
        \caption{GACT alignment with $(T, O) = (32, 7)$.}
        \label{fig:single_pass_more_overlap_not_always_better_large}
    \end{subfigure}
    \caption{Case in which GACT, with a smaller overlap configuration, does better than GACT with a larger one. Having a smaller overlap configuration means that neighboring tiles are placed further apart and can reach certain high-scoring cells that would not be attained at larger overlap values.
    Green sections of the traceback represent areas where GACT's traceback matches that of the optimal alignment, and magenta regions show the remaining traceback of the optimal alignment.
    The circle at the tip of the tracebacks signal the presence of a zero cell that causes the procedure to stop.}
    \label{fig:single_pass_more_overlap_not_always_better}
\end{figure*}

Figure~\ref{fig:single_pass_more_overlap_not_always_better} shows one such example where an optimal alignment is achieved with $(T, O) = (32, 6)$, but where the traceback ends prematurely with $(T, O) = (32, 7)$.
This phenomenon can be explained by analyzing the values of the upper-left cells encompassed by the 2nd tile of Figure~\ref{fig:single_pass_more_overlap_not_always_better_small} that are absent in the same tile of Figure~\ref{fig:single_pass_more_overlap_not_always_better_large}.
The highly variable values in the substitution matrix, coupled with the relatively low scores in GACT tiles, means that displacing a tile by just 1 cell can cause it to miss a high-scoring substitution, and therefore trigger a sharp drop in tile cell values that ultimately cause a traceback to end prematurely at a zero cell.

The conclusion is that a GACT suffers excessively for small local alignment mistakes on protein sequences, a situation that can also occur for DNA sequences, but which is far less probable due to the low amplitude scores in DNA substitution matrices that smooths error propagation.

\subsection{Hardware resource requirements}
The impressive alignment throughput of Darwin relies on two mechanisms: (1) each GACT module is an array of PEs that compute S\&W cells in a wavefront progression; and (2) GACT modules are replicated 64 times to handle incoming seed hits provided by D-SOFT in parallel.

The traditional systolic architecture used to implement S\&W is economical in hardware resources when used for DNA sequences as the substitution ``matrix'' can simply be replaced with a comparator, two registers to store the match reward or mismatch/gap penalty, and finally a multiplexor that selects one of the two values.

This optimized implementation cannot be used when protein sequences are being aligned, as the substitution matrix is too large to be modeled with simple logic/registers and must be stored in an on-chip memory per-PE for fast lookup.
It is therefore not possible to put as many PEs in each GACT module given the increased on-chip memory requirement, and thus processing large tiles becomes linearly slower for protein sequences.

\section{Improving the GACT heuristic}
\label{sec:multi_pass_gact}

Based on our observations, we need to reduce:

\begin{itemize}
    \item[--] Premature traceback termination due to sensitivity to GACT's placement of the \emph{current} tile;
    \item[--] Traceback divergence due to an erroneous traceback in a \emph{previous} tile that causes the \emph{current} tile from being misplaced.
\end{itemize}

The last point stems from the fact that GACT uses its current traceback to help place the \emph{next} tile, but this assumes that the traceback leading up to the tile borders is correct.
This assumption is one of the causes of ``forward'' traceback divergence, but we show that the phenomenon can be attenuated with some additional computation.

All issues with GACT arise from low scores being present in the S\&W scoring matrix.
A natural solution to low-scoring cells would be to increase their values when we detect they are falling too low, however we cannot perform such an operation since one does not know what ``too low'' is\footnote{Long alignments would have a high ``low threshold'', whereas short alignments would have a low ``low threshold'', but we do not know in advance whether the alignment is long or short.}.
Instead we propose the following 6-step multi-pass algorithm that increases confidence in the placement of the \emph{next} tile by increasing confidence that the traceback of the \emph{previous} tile is correct, while simultaneously retaining GACT's key contribution (constant traceback memory usage):

\begin{enumerate}
    \item Compute cells of tile~1;
    \item Perform traceback of tile~1 until distance $O$ from border;
    \item Compute cells of tile~2 (no traceback);
    \item Recompute cells of tile~1, taking into account elevated values from overlapped region with tile~2;
    \item Perform final traceback of tile~1 until distance $O$ from border;
    \item Advance to the next tile.
\end{enumerate}

\begin{figure*}
    \centering
    \begin{subfigure}{.45\textwidth}
        \centering
        \includegraphics[width=0.8\linewidth]{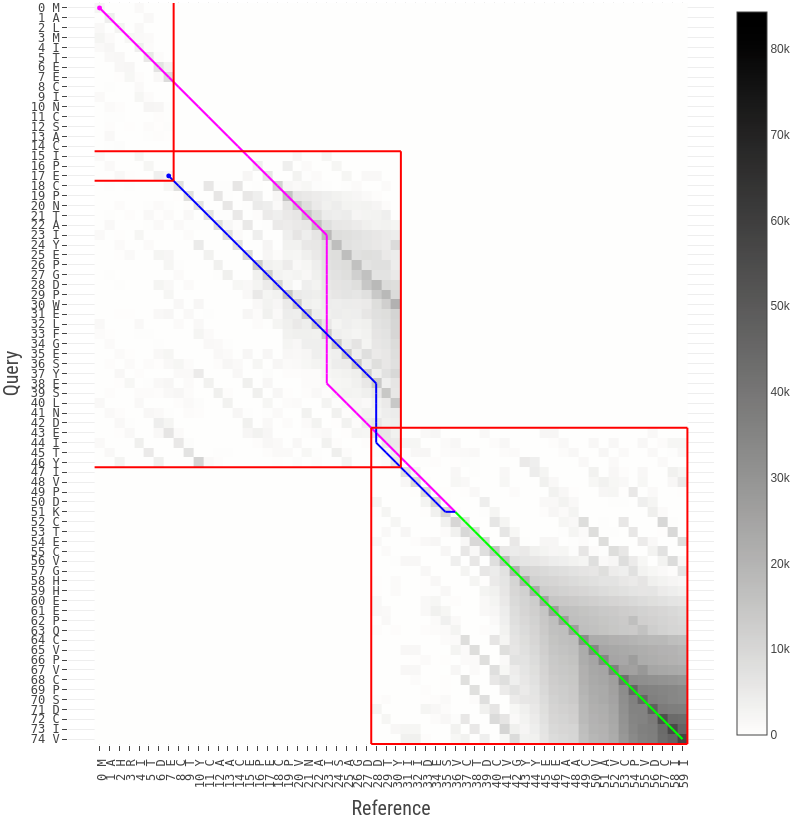}
        \caption{Single-pass GACT alignment with $(T, O) = (32, 3)$.}
        \label{fig:single_pass}
    \end{subfigure}%
    \begin{subfigure}{.45\textwidth}
        \centering
        \includegraphics[width=0.8\linewidth]{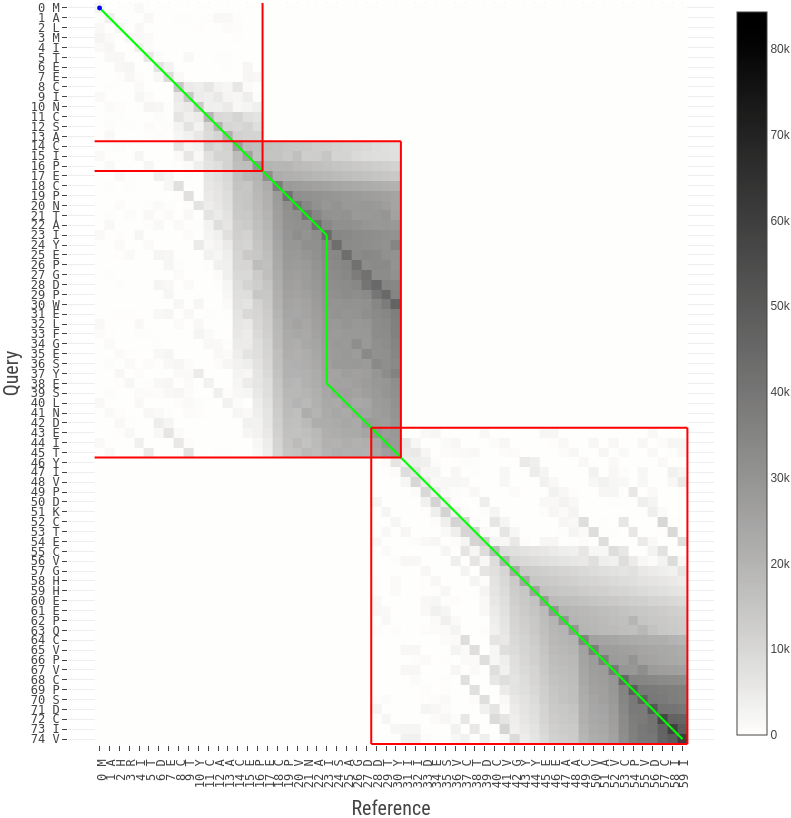}
        \caption{Multi-pass GACT alignment with $(T, O) = (32, 3)$.}
        \label{fig:multi_pass}
    \end{subfigure}
    \caption{Single-pass vs. multi-pass GACT heuristic alignment.
    Notice how the multi-pass variant of the algorithm brings out significantly higher cell values in the second tile due to judicious placement of the third tile that now passes through a high-scoring region of the optimal alignment.
    The blue line represents the traceback obtained by GACT, the magenta line represents the traceback of the optimal alignment, finally the green line marks regions where both GACT and the optimal alignment match with equality.}
    \label{fig:single_vs_multi_pass}
\end{figure*}

Figure~\ref{fig:single_pass} shows an example where the original GACT algorithm suffers from an erroneous traceback in a \emph{previous} tile, which then causes a future tile to be misplaced (by just 1 cell) and ultimately lead to premature traceback termination.
In turn, Figure~\ref{fig:multi_pass} shows how the proposed multi-pass variant of GACT handles such a situation by raising cell scores enough such that new traceback ``routes'' that were previously unseen become available.

Note that the algorithm is completely determined by the dataset on which it operates and does not rely on any pre-determined ``low threshold'' for cell boosting.
Furthermore, one does not need to recompute the values of cells that were overlapped with tile~2 as they can be forwarded from tile~2's on-chip memory.

When evaluated on a small bacterial protein dataset containing roughly 25K sequences, we found that our multi-pass variant obtains on average 14\% better alignment scores when used under identical parameters as the original GACT algorithm, with relative score improvements ranging from 0\% (for identical alignments as GACT) up to 29000\% (for alignments where the multi-pass variant allows long sequences to align much better).

Note that tile-based alignment can be accelerated if (1)~tiles are smaller, i.e. quadratically fewer cells to compute, and (2) tiles are laid out as far as possible from each other, reducing cell value recomputation.
Though it is true that this multi-pass variant of GACT roughly triples the compute requirements of the original algorithm in terms of number of tiles computed, we believe that the increased confidence in the tracebacks produced allows us to minimize tile sizes and overlap margins while obtaining similar results, therefore enabling a quadratic reduction in the number of cells that need to be computed.

Ultimately, GACT remains a heuristic algorithm and while our proposed enhancement alleviates some of its issues, we do not eliminate them.
We discuss possible solutions in Section~\ref{sec:interval_alignment}.

\section{Future Work}
\label{sec:interval_alignment}

\subsection{Problematic sequences}
One notable issue that neither the original GACT algorithm nor our multi-pass variant solves is that of traceback diverging due to the presence of a single long deletion or insertion in the optimal alignment.
Indeed, it is impossible for any tile-based alignment algorithm to find the optimal traceback if a deletion or insertion has a length longer than the chosen tile size as scores would obviously reach zero.
Furthermore, increasing overlap margins is no solution here as successively placing a tile where it goes through such ``vertical'' or ``horizontal'' traceback strips will ultimately cause the tile to no longer cover the region of interest as tiles are always placed diagonally from each other by at least one cell.
The only way for a tiled alignment to work on such problematic sequences is if the tile size is larger than the longest deletion or insertion, but one cannot know this value in advance.
Increasing tile sizes also has a negative effect on alignment speed.

GACT and our multi-pass variant are therefore well-suited for finding ``diagonal'' alignments in a S\&W score matrix, but not for long ``vertical'' or ``horizontal'' ones.
This brings us to our discussion of D-SOFT, Darwin's filtering algorithm, and how we can improve it to help solve the long deletion/insertion problem.

\subsection{Interval-based alignment}
Myers' proposed an algorithm for finding seed hits that have a high probability of being ``close'' to the optimal alignment in the presence of noisy long reads~\cite{Myers:2014:IPA:978-3-662-44753-6_5}.
Roughly, the idea is to (1) find \emph{exact} matches of seeds of size $k$ from the query in the reference, then (2) count the number of non-overlapping seed hits in diagonal bands, and finally (3) select a band that passes a certain threshold.
Darwin's D-SOFT implementation translates Myers' hardware-unfriendly algorithm into one more suitable for hardware acceleration and is able to provide quality seeds to the GACT arrays for extension.

However, D-SOFT is unsuitable for use with protein sequences for one main reason:
While the probability of finding sequences of 10-15 identical nucleotides in long DNA sequences is very high, the probability of finding a sequence of even three common amino-acids between two proteins is almost zero.
Therefore any seeding algorithm that relies on exact matches to find seeds will experience seed starvation.


Since exact string matching does not work, we envision a filtering algorithm that uses the same subsequence matching method as S\&W.
Concretely, we propose to select multiple small samples from the query and the reference sequences, and to perform independent S\&W computations on these regions to obtain a single value for each, effectively that of the max-scoring cell.
We can then find diagonal sequences of high-scoring ``blocks'' in this ``down-sampled'' space and issue them as seeds to the GACT arrays for alignment.
Finally, these aligned segments would need to be stitched together to form a complete alignment without affecting their alignment score.

What still remains to be determined is how big the sampled blocks must be and how this decision affects the choice of tile sizes that GACT can use afterwards to guarantee it can handle all (smaller) vertical/horizontal jumps that may be encountered during its alignments.

\section{Conclusion}
\label{sec:conclusion}

This paper analyzed the limitations of Darwin's design for protein sequence alignment, a more difficult problem due to the increased search space and hardware requirements.
It proposes a multi-pass aligner variant that helps increase the aligner accuracy of Darwin on protein sequences.
Our proposed multi-pass variant of GACT achieves on average 14\% better alignment scores.
We also discuss a refinement to Darwin's filtering algorithm that can be used in conjunction with our multi-pass algorithm to bridge a gap that is currently yet unhandled by both Darwin's aligner and our multi-pass variant: obtaining good alignments under the presence of large insertions and deletions.

\section*{Acknowledgements}
We thank Yatish Turakhia for sharing the code for his D-SOFT filter and GACT aligner upon which this work is based, as well as for his valuable assistance.

\clearpage


\bibliographystyle{ieeetr}
\bibliography{ref}

\end{document}